\pdfoutput=1
\documentclass[12pt,letter]{article}

\usepackage[utf8]{inputenc}

\usepackage{graphicx}
\usepackage{microtype}
\usepackage{csquotes}
\usepackage{siunitx}
\usepackage{hyperref}
\hypersetup{colorlinks=true,allcolors=black}
\usepackage{cleveref}
\usepackage{amsmath}
\usepackage{amssymb}
\usepackage[square,sort&compress,comma,numbers]{natbib}
\usepackage[parfill]{parskip}

\usepackage{scalefnt}
\newcommand{\abbrev}{\scalefont{.9}}

\newcommand{\NNNLO}{\text{\abbrev N$^3$LO}}
\newcommand{\NNLO}{\text{\abbrev NNLO}}
\newcommand{\NLO}{\text{\abbrev NLO}}
\newcommand{\LO}{\text{\abbrev LO}}
\newcommand{\EFT}{\text{\abbrev EFT}}
\newcommand{\SCET}{\text{\abbrev SCET}}

\newcommand{\SM}{\text{\abbrev SM}}

\newcommand{\QCD}{\text{\abbrev QCD}}
\newcommand{\EW}{\text{\abbrev EW}}
\newcommand{\PDF}{\text{\abbrev PDF}}

\newcommand{\CP}{\text{\abbrev CP}}
\newcommand{\LHC}{\text{\abbrev LHC}}

\newcommand{\MCFM}{\text{\abbrev MCFM}}
\newcommand{\ATLAS}{\text{\abbrev ATLAS}}
\newcommand{\CMS}{\text{\abbrev CMS}}

\newcommand\pubnumber{CIPANP2018-Neumann, IIT-CAPP-18-03, FERMILAB-CONF-18-539-T}
\newcommand\pubdate{October 3, 2018}

\def\Title#1{\begin{center} {\Large #1 } \end{center}}
\def\Author#1{\begin{center}{ \sc #1} \end{center}}
\def\Address#1{\begin{center}{ \it #1} \end{center}}

\newcommand\pubblock{\rightline{\begin{tabular}{l} \pubnumber\\
         \pubdate  \end{tabular}}}
\newenvironment{Abstract}{\begin{quotation}  }{\end{quotation}}
\newenvironment{Presented}{\begin{quotation} \begin{center} 
             PRESENTED AT\end{center}\bigskip 
      \begin{center}\begin{large}}{\end{large}\end{center} \end{quotation}}

\begin{document}
\begin{titlepage}
\pubblock

\vfill
\Title{Recent Developments in Gluon Fusion Higgs Calculations}
\vfill
\Author{Tobias Neumann}
\Address{
	Department of Physics, Illinois Institute of Technology, Chicago, Illinois 60616, USA\\
	Fermilab, PO Box 500, Batavia, Illinois 60510, USA
	}
\vfill
\begin{Abstract}
During recent years perturbative fixed order and resummation calculations have decreased uncertainties on predictions 
for gluon fusion 
Higgs production cross sections tremendously. Most exciting results have been 
published just this year. In these proceedings I present an overview of recent and most recent developments of these 
calculations that allow theory predictions to compete with the experimental precision reached by future collider 
upgrades.
\end{Abstract}
\vfill
\begin{Presented}
Thirteenth Conference on the Intersections of\\ Particle and Nuclear Physics (CIPANP 2018)\\~\\
Indian Wells, California, USA, May 29 to June 3, 2018
\end{Presented}
\vfill
\end{titlepage}
\def\thefootnote{\fnsymbol{footnote}}
\setcounter{footnote}{0}

\subsection*{Introduction}

Over the run-time of the \LHC{} and its associated experiments, many people's expectations in terms of reached 
precision have been surpassed. 
Although true percent level or sub-percent level precision physics is not within its scope, even the currently reached 
and near future reached precision poses a challenge for the theory community calculating Standard Model (\SM{}) 
predictions. This is especially true for gluon fusion Higgs 
production, which is one of the cornerstone Higgs production mechanisms by virtue of its large cross section compared 
to the other channels.

Recent experimental results from \ATLAS{} and \CMS{} predominated by the gluon fusion production mechanism are widely 
about to 
crack the 
10\% level in statistical uncertainties 
\cite{Aaboud:2018jqu,Aaboud:2018puo,Aaboud:2018ezd,Aaboud:2018xdt,Aaboud:2017vzb}, 
\cite{CMS:2018lkl,Sirunyan:2018kta,Sirunyan:2018egh}. Broadly speaking, provided that methods of data analysis and 
calculational techniques advance as expected, we can expect single digit uncertainties for the inclusive gluon fusion 
Higgs cross section and about $10\%$ uncertainties on differential quantities at a $\SI{3000}{fb^{-1}}$ high 
luminosity upgraded \LHC{} \cite{ATL-PHYS-PUB-2014-016}. A large part of that burden is carried by the uncertainties of 
theory predictions which ideally should be significantly smaller than experimental statistical and systematic 
uncertainties.

An important result this year for experiment and theory is the observation of the $H\to b\bar{b}$ decay 
\cite{Sirunyan:2018kst} through the combination of many individual analysis channels. The gluon fusion production 
mechanism only plays a small role in this result due to large experimental, but also large theory uncertainties 
\cite{Sirunyan:2017dgc}. Its analysis was performed for a highly boosted Higgs boson and relied on Higgs+jet 
predictions which at that time had large uncertainties associated with the top-quark mass. This uncertainty has been 
mostly eliminated meanwhile as I will also highlight further below. 

Although pushing gluon fusion Higgs production uncertainties to the percent level will be a task of future \LHC{} 
upgrades and future colliders, theory predictions should match experimental ones, and be ideally significantly 
lower. The current goal is thus to provide theory predictions at the single percent level uncertainty, which is an 
extraordinarily difficult task given the many ingredients that enter a full Standard Model calculation at a hadron 
collider. The low hanging fruits are gone and it requires the effort of larger and larger groups and 
collaborations to stem this problem.

\subsection*{Reach for precision}

The prediction of the inclusive gluon fusion Higgs production cross section made a big step forward a few years ago 
through the calculation of the \NNNLO{} \QCD{} corrections \cite{Anastasiou:2015ema}. This was made possible by using 
the limit of an infinitely heavy top-quark (\EFT{}) and an expansion around the Higgs production threshold.
An assessment of all uncertainties was soon-after performed \cite{Anastasiou:2016cez}, estimating the residual theory 
uncertainty and uncertainties in \PDF{}s and $\alpha_s$ to about less than $10\%$. This estimate is based on the
aforementioned \NNNLO{} calculation in the \EFT{} approximation augmented with finite top-quark and bottom-quark mass 
effects, threshold resummation and mixed QCD-electroweak correction factors.

For differential gluon fusion Higgs production with a recoiling jet, \NNLO{} 
predictions in 
the \EFT{} have uncertainties of $10\%$ just from the truncation of higher order \QCD{} effects, which has been known 
for some time now \cite{Boughezal:2013uia,Chen:2014gva,Boughezal:2015aha}. Of high importance is the Higgs 
transverse momentum $(p_T)$ distribution, especially in the boosted high $p_T$ regime. Further effects with 
associated uncertainties like transverse momentum resummation, electroweak corrections and finite bottom and top-quark 
masses need to be taken into account and increase 
the uncertainties each. At this point \PDF{} and $\alpha_s$ uncertainties were not even mentioned yet. For example in 
the case of the 
gluon fusion $H\to b\bar{b}$ analysis \cite{Sirunyan:2017dgc}, an uncertainty of $30\%$ is assigned to the theory 
prediction.

Going from these stacks of uncertainties in inclusive and differential Higgs production to a combined one percent 
uncertainty is a task that at the very least must be reached by each individual ingredient. At the very least we want 
to sufficiently 
surpass the current and near-future experimental precision. The consistent combination of pieces to truly reach one 
percent uncertainty seems 
like a task that is too difficult to even think about. Yet the theory community must work towards this goal and has 
made important progress just this year, of which I will give a summarizing excerpt of most important results in the 
following paragraphs.

I will limit myself to present only recent results of Standard Model gluon fusion Higgs production in the inclusive and 
differential case. I will also mostly just refer to studies presenting phenomenological results, and not to the equally 
important long tail of technical publications that paved the road. For further results see for example the recent 
Standard Model 
working group report \cite{Bendavid:2018nar} and a most recent review article on Higgs physics \cite{Dawson:2018dcd}.

\subsection*{Recent developments}

A direct improvement of the inclusive gluon fusion Higgs cross section has been achieved through an exact calculation 
at \NNNLO{} in the \EFT{} removing the need for a threshold expansion \cite{Mistlberger:2018etf}. While the associated 
threshold expansion uncertainty was estimated to be $\simeq0.5\%$ and turned out to be a little less, it is an 
important step not just for the reduction of the uncertainty, but also for the experience gained from the occurrence of 
elliptic integrals. More specifically, dedicated progress has been made this year to classify elliptic polylogarithms 
and to develop them to the same degree as non-elliptic polylogarithms 
\cite{Broedel:2018iwv,Broedel:2018qkq,Broedel:2018qkq}.

\begin{figure}
	\centering\includegraphics[]{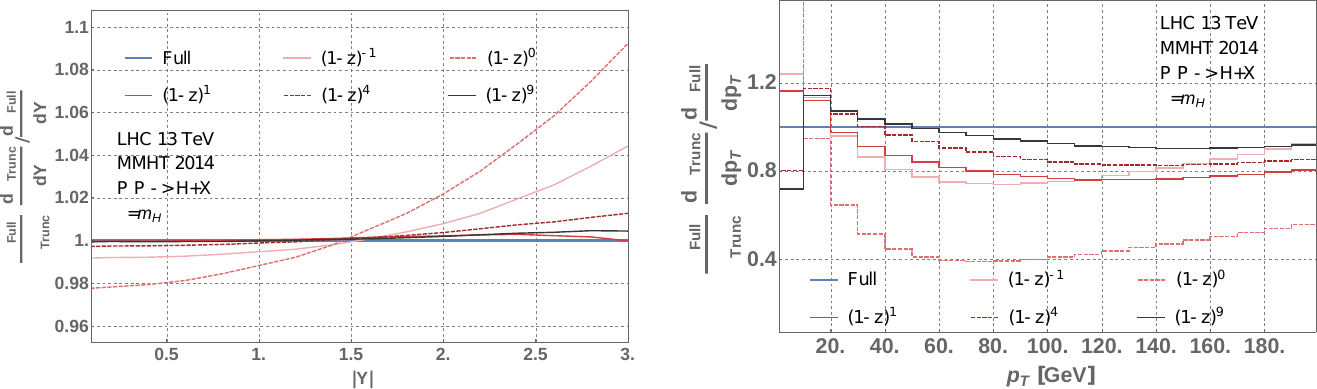}
	\caption{Normalized \NNLO{} rapidity (left) and transverse momentum (right) distributions for the Higgs boson in 
	the threshold expansion $(1-x)^n$ and as a full result. The 
		normalization and full result is given as the blue line. Higher orders in the threshold expansion are 
		reweighted such that their cumulant reproduces the unexpanded \NNLO{} cross section. Plot taken from 
		ref.~\cite{Dulat:2017prg}.}
	\label{fig:Dulat:2017prg}
\end{figure}

Using the threshold expansion for Higgs differential distributions has been followed in 
ref.~\cite{Dulat:2017prg}. It was found that in principle a threshold expansion is sufficient to determine the 
analytic Higgs rapidity distribution at N$^3$LO at the few percent level. This was estimated from a comparison 
between 
the threshold expansion and exact results both at \NNLO{}. Less 
inclusive quantities like transverse momentum can not be approximated sufficiently by the expansion, see 
\cref{fig:Dulat:2017prg}. Even if not used 
directly, the threshold expansion can provide valuable boundary information for the computation of master integrals 
using differential equations.

\begin{figure}
	\centering\includegraphics[width=0.5\textwidth]{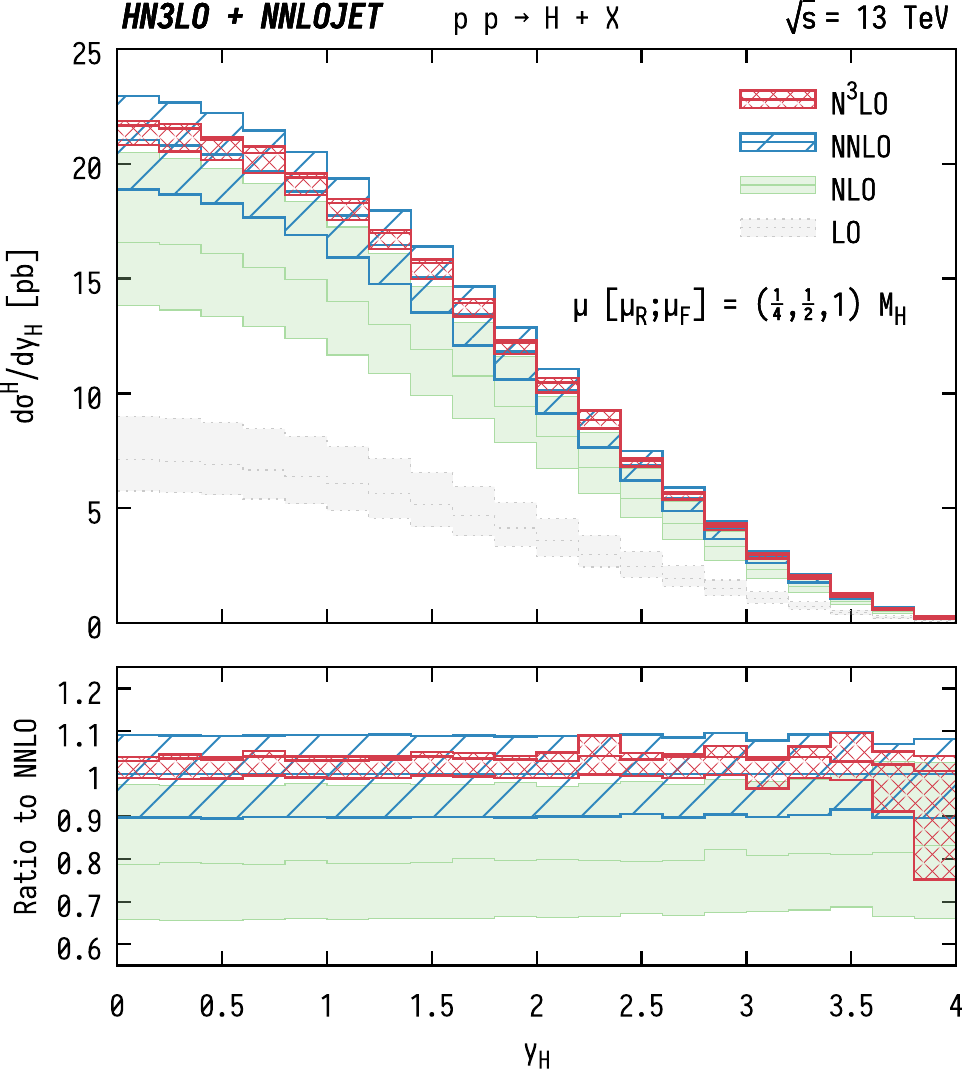}
	\caption{Higgs rapidity distribution up to N$^3$LO with scale variation uncertainties. At N$^3$LO also the 
	$q_T^\text{cut}$ variation uncertainty and uncertainty from the numerical extraction of a $\delta(q_T^2)$ piece is 
	taken into account. For details see \cite{Cieri:2018oms}, where this plot is taken from.}
	\label{fig:rapn3lo}
\end{figure}

The \NNNLO{} cross section has meanwhile also been obtained in the $q_T$ subtraction 
formalism \cite{Cieri:2018oms}. This result relies on the previously calculated \NNNLO{} cross section though to match 
and 
numerically extract a $\delta(q_T^2)$ piece. Using this, the authors were able to obtain the Higgs rapidity 
distribution at \NNNLO{}, which comes with a significant reduction of scale uncertainty compared to \NNLO{}, see 
\cref{fig:rapn3lo}.

A public implementation  of inclusive gluon fusion Higgs production at \NNNLO{} is available with the code iHixs
\cite{Dulat:2018rbf} which also includes some effects beyond the fixed order \EFT{} result. The competing code 
SusHi, while still relying on the threshold expansion and a matching to the high energy limit, is more focused 
towards physics beyond the Standard Model as it can also provide predictions for a \CP{} odd scalar and dimension five 
Higgs-gluon coupling 
operators.

\begin{figure}
	\centering\includegraphics[]{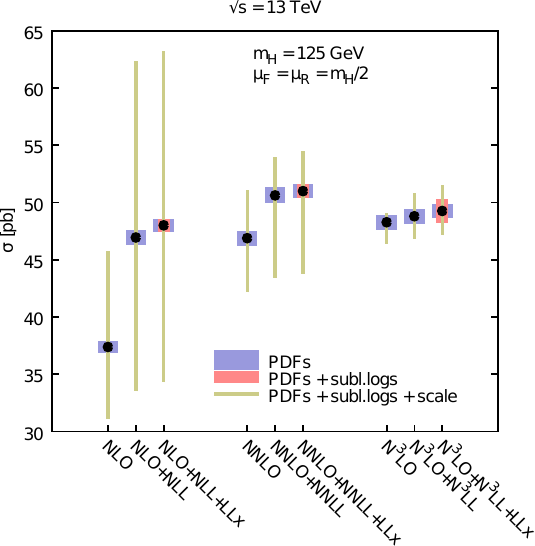}
	\caption{Inclusive gluon fusion Higgs cross sections at fixed orders \NLO{}, \NNLO{} and \NNNLO{} with added 
	threshold resummation and double resummation. For details regarding the uncertainties see 
	ref.~\cite{Bonvini:2018ixe}, where this plot is taken from.}
	\label{fig:doubleresum}
\end{figure}

The first double resummed prediction for inclusive gluon fusion Higgs production is presented in 
ref.~\cite{Bonvini:2018ixe}. Both logarithmic corrections at production threshold and in the high energy limit are 
simultaneously resummed, the former through {\abbrev N$^3$LL}, the latter at {\abbrev LL}, and matched to the \NNNLO{} 
fixed order result. Additionally the effect of resummed \PDF{}s is included \cite{Ball:2017otu}. The implementation is 
based on publicly available codes for threshold and high-energy resummation, and includes some top-quark mass effects. 
The authors find a correction of $2\%$ on the fixed order cross section and provide most recent and accurate cross 
section predictions. It is important to note that this calculation opened up a new source of uncertainty, and that as 
such the total uncertainty increases due to missing subleading high energy logarithms, see \cref{fig:doubleresum}.

Progress has also been made in re-evaluating the size of electroweak effects in gluon fusion 
\cite{Bonetti:2018ukf,Bonetti:2017ovy}. Electroweak (\EW{}) effects enhance the pure \LO{} \QCD{} 
Higgs cross section by about $5\%$ \cite{Aglietti:2004nj,Actis:2008ug}. A calculation of electroweak corrections at 
\NLO{} 
in \QCD{} at the three-loop level is important to assess the factorization approximation of \QCD{} $\times$ electroweak 
corrections. This allows one to apply \EW{} correction factors to higher order \QCD{} cross sections. Previous results 
estimated those \EW{} corrections in an approximation valid for $m_H < m_W$  
\cite{Anastasiou:2008tj} and found an equal enhancement of about $5\%$ with respect to the \NLO{} \QCD{} result. This 
clearly
supports the \QCD{} $\times$ \EW{} factorization. The new study presented in ref.~\cite{Bonetti:2018ukf} employs a
soft-gluon approximation for the real emission contributions but otherwise takes analytical results for the three loop 
virtual corrections to confirm those earlier estimates.
  
\paragraph{Higgs in association with a jet.} I will now focus on improvements on the Higgs+jet cross sections, where 
the transverse momentum ($p_T$) distribution of the Higgs boson is of particular interest. Soft gluon resummation has 
to be accounted for at low $p_T$ as well as light quark masses, while at large $p_T$ top-quark mass effects play the 
biggest role.

\begin{figure}
	\centering\includegraphics[]{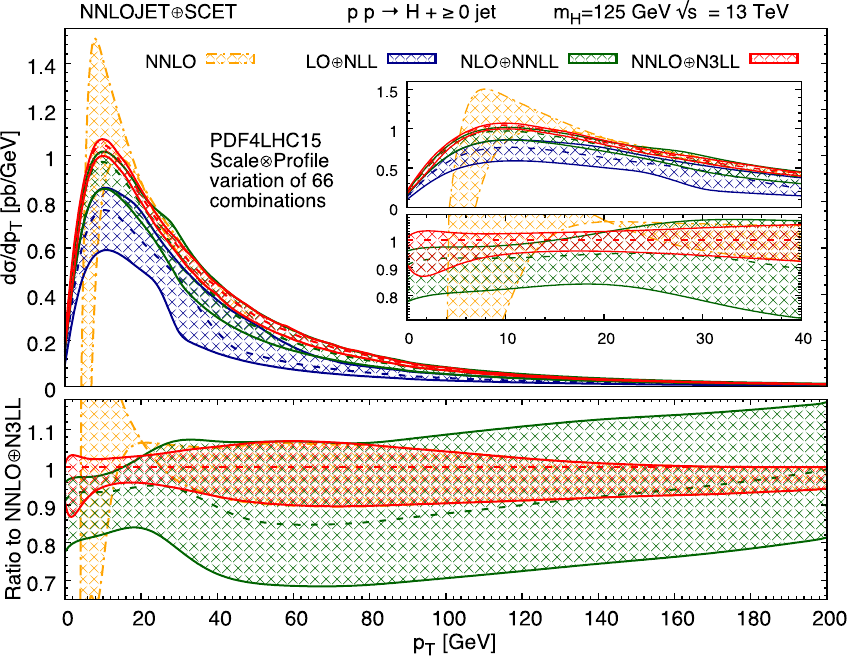}
	\caption{Higgs transverse momentum distribution at different orders with a central renormalization scale of $m_H/2$ 
		and a resummation matching at \SI{30}{\GeV}. The uncertainties are obtained by scale variation and varying the 
		resummation matching. Taken from ref.~\cite{Chen:2018pzu}.}
	\label{fig:ptresum}
\end{figure}

Transverse momentum resummation for Higgs+jet production at \NNLO{} has meanwhile been performed at N$^3$LL on top of 
the \NNLO{} fixed order \EFT{} result. This has been achieved in two ways. It has first been calculated in a 
Monte-Carlo 
approach to resum directly in $p_T$ space \cite{Bizon:2017rah,Bizon:2018foh} and has then been evaluated analytically 
for an 
improved resolution in a \SCET{} framework \cite{Chen:2018pzu}. The analytically resummed $p_T$ distribution at 
different perturbative orders is shown in \cref{fig:ptresum}.

\begin{figure}
	\centering\includegraphics[]{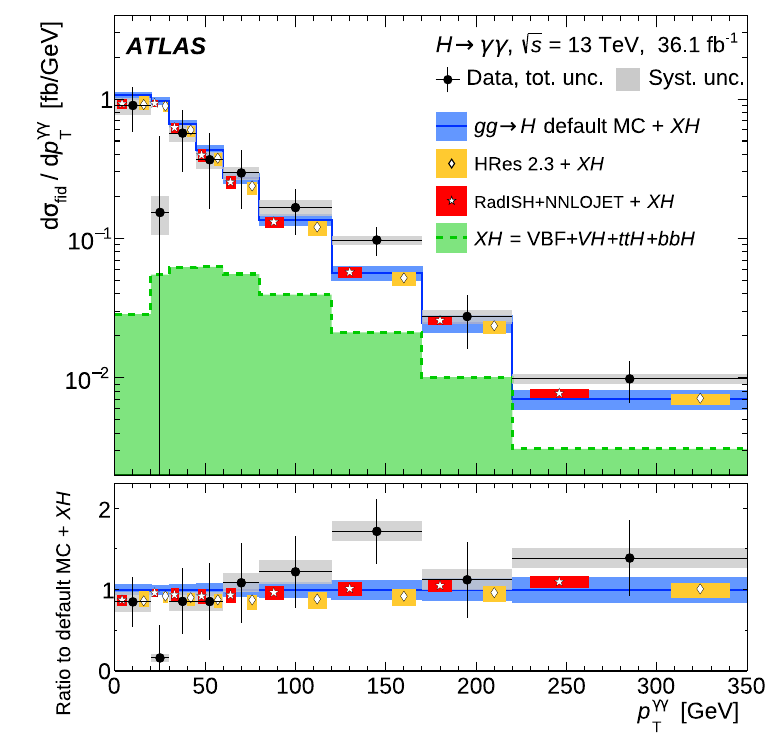}
	\caption{Measured differential gluon fusion Higgs transverse momentum distribution ($p_T$ of the two photons) at 
	the \ATLAS{} 
	experiment. Figure taken from ref.~\cite{Aaboud:2018xdt}. }
	\label{fig:ptgamgam}
\end{figure}

The intermediate transverse momentum region $m_b \lesssim p_T \lesssim m_T$ is of particular importance for current and 
previous experimental analyses that have no precision reach at high $p_T$ yet. See for example \cref{fig:ptgamgam}, 
where the $p_T$ distribution of the $H\to\gamma\gamma$ system as measured by \ATLAS{} is shown. The large number of 
mass scales in this region makes it theory-wise difficult to compute it 
precisely. Logarithms of the kinematic ratios $p_T / m_b$ and $m_H / m_b$ can grow large, and their all-order 
resummation is currently out of reach. 

Soft gluon resummation and bottom quark mass effects, which are most 
important in that region, were taken into account as far as currently possible in 
ref.~\cite{Caola:2018zye}. The inclusion of {\abbrev NNLL} soft gluon resummation and bottom quark mass effects lead to 
an increase of $\simeq 5\%$ of the \NLO{} cross section. The study focuses on assessing 
uncertainties in this important region and finds that the uncertainty on the predominant top-bottom interference has a  
$\simeq 15\%$ scale uncertainty. An additional uncertainty of up to $\simeq 20\%$ is due to the bottom quark mass 
renormalization scheme at small $p_T$. Combining these with the uncertainty that comes from ambiguities in the soft 
gluon resummation in presence of the massive bottom quark, they conclude that generally uncertainties are about $\simeq 
20\%$ in the discussed region. As such, one can expect further percent level effects from bottom quark mass effects in 
this intermediate energy region.

Finally, the region of large transverse momenta, becoming highly relevant through boosted Higgs analyses 
\cite{Sirunyan:2017dgc} has received a large improvement by removing the top-quark mass uncertainty at \NLO{}. This 
uncertainty 
is due to the \EFT{} approximation, reducing the number of loops to compute not just by one, but also removing the 
top-quark mass itself in the loop integrals. It is only valid in the limit $p_T \ll 2 m_t$. Using the two-loop virtual 
contributions in a large energy expansion \cite{Neumann:2018bsx,Lindert:2018iug}, and through a numerical evaluation of 
the exact integrals by sector decomposition \cite{Jones:2018hbb}, the top-quark mass uncertainty was eliminated at 
\NLO{}.

\begin{figure}
	\centering\includegraphics[]{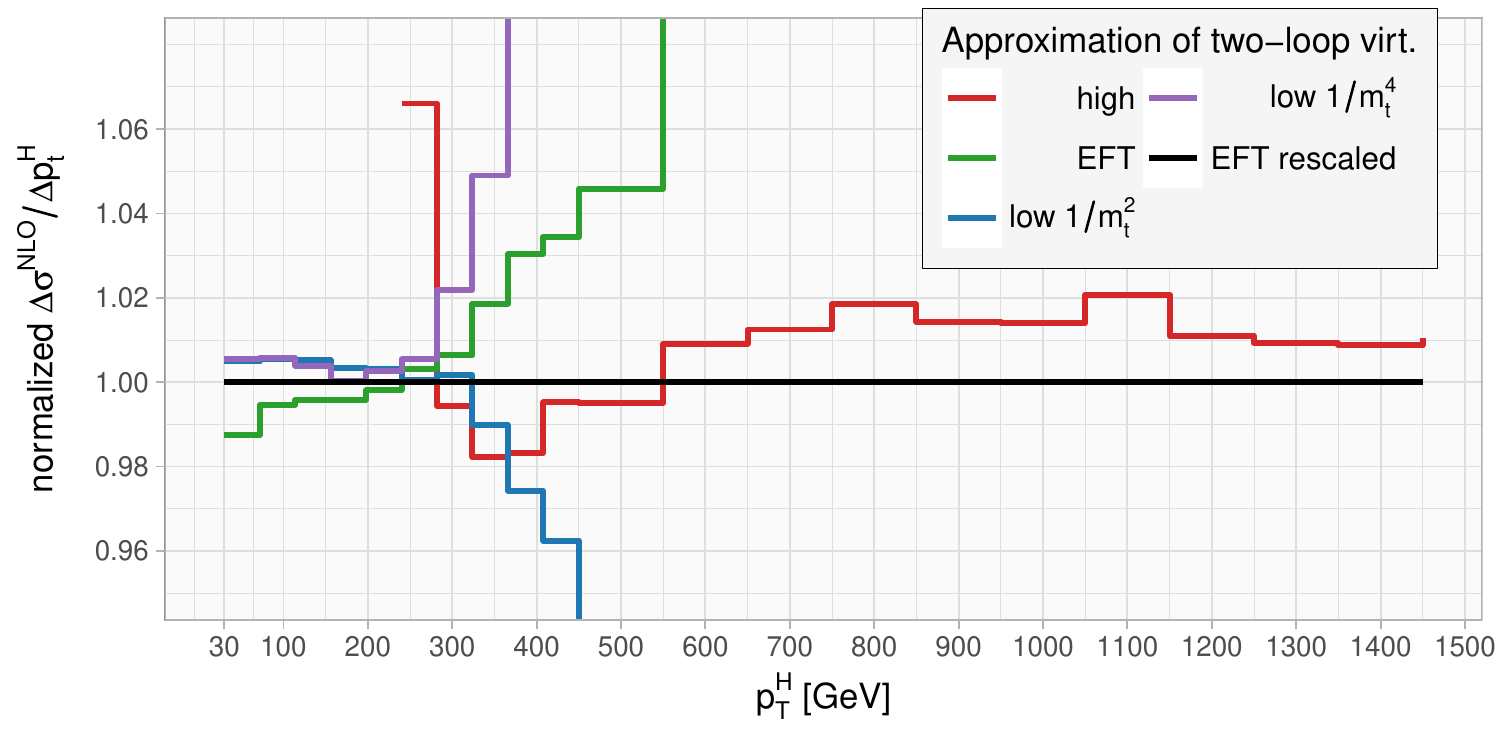}
	\caption{Normalized \NLO{} Higgs transverse momentum distribution with top-mass dependence in various 
	approximations. The normalization (EFT rescaled) denotes the approximation where only the finite part of two-loop 
	virtual 
	correction in the \EFT{} approximation	is point wise rescaled by the full top-mass dependent one-loop amplitude; 
	real emission and born part are exact in $m_t$. Using the high energy approximation for the missing two-loop 
	virtual	amplitudes results in the red line, while the other lines denote low energy approximations. For 
		details see ref.~\cite{Neumann:2018bsx}, where this plot is taken from.}
	\label{fig:pthhpt}
\end{figure}

\begin{figure}
	\centering\includegraphics[]{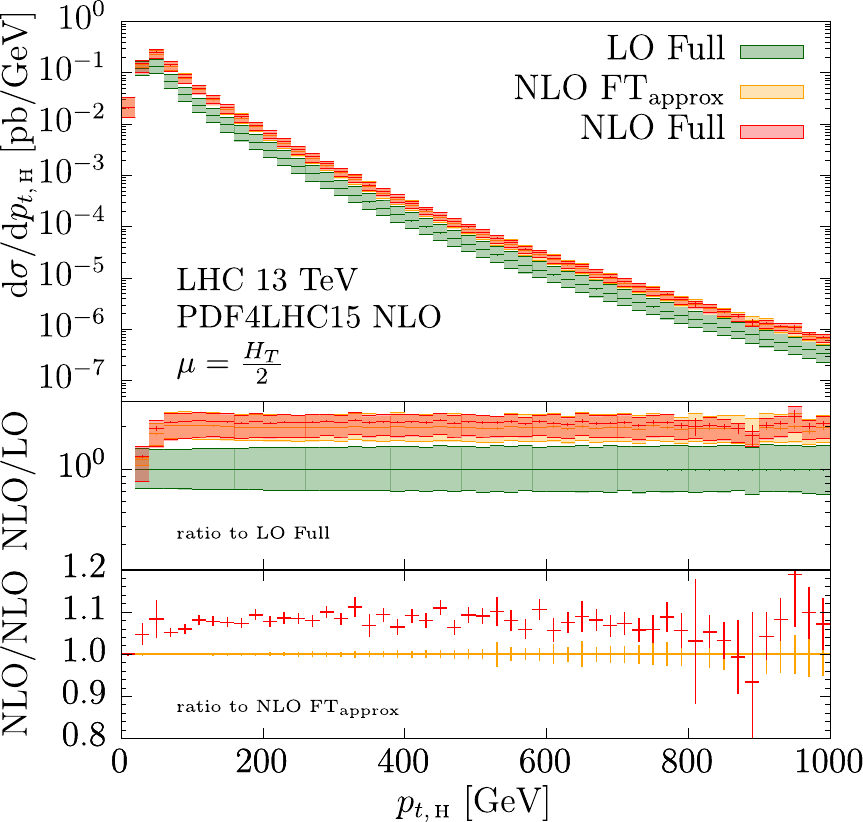}
	\caption{\NLO{} Higgs transverse momentum distribution with exact top-mass dependence in 
		comparison to an approximation that uses rescaled virtual corrections (FT$_\text{approx}$). For details see 
		ref.~\cite{Jones:2018hbb}, 
		where this plot is taken from.}
	\label{fig:pthsd}
\end{figure}

The corrections due to a finite mass top-quark in the virtual corrections are $\simeq 1-2\%$ for $p_T \gg m_T$, when 
compared to an approximation with rescaled virtual corrections, see \cref{fig:pthhpt}. A similar comparison is 
performed in ref.~\cite{Jones:2018hbb} using the numerical evaluation of the exact top-mass dependent integrals. 
Relatively flat corrections of roughly $5\%$ over the whole $p_T$ spectrum are found, see \cref{fig:pthsd}.

Both the high energy approximated two-loop amplitude and the 
numerical evaluation of the full two-loop amplitude lead to compatible results that remove the previously unconstrained 
top-quark mass uncertainty at \NLO{} at large transverse momenta. The high energy approximate results published in 
ref.~\cite{Neumann:2018bsx} are publicly available through the code \MCFM{}-8.2 and are combined with an implementation 
of the analytical top-quark mass exact real emission amplitudes \cite{Neumann:2016dny}.

As of now no analytical expressions of the two-loop amplitudes necessary for the exact top-mass dependent virtual 
corrections of Higgs+jet at \NLO{} are available. Analytical results for the necessary two-loop planar integrals in the 
euclidean 
region have been published a while ago \cite{Bonciani:2016qxi}. Solutions for the non-planar integrals are not yet 
available but promised to be ready soon\footnote{See talk by Hjalte Frellesvig at Loops \& Legs 2018.}. On the 
forefront of efficient analytical one loop 
amplitudes progress has been made for Higgs + $n$-gluons. For $n\leq5$ very compact expressions have been 
derived in ref.~\cite{Ellis:2018hst} when all gluons have the same helicity. Extending this to the more complicated 
helicity combinations would offer the prospect of a very fast and stable numerical evaluation. This is highly 
important, as the high dimensional real emission integration usually takes most of the time to calculate cross 
sections.

Top-mass effects in Higgs+jet production have traditionally been taken into account at low $p_T$ through a $1/m_t^n$ 
expansion at \NLO{}, valid at energies $p_T\lesssim 2 m_T$, see for example ref.~\cite{Neumann:2016dny}. The 
computation of 
analytic two-loop amplitudes including dimension $7$ Higgs-gluon operators allows to include such effects at \NNLO{} 
\cite{Jin:2018fak,Brandhuber:2018xzk,Brandhuber:2018kqb}.

Finally, advancements are not just made in calculating cross sections, but also to analyze the data. Regarding this, 
modern 
machine learning and jet substructure techniques are being developed \cite{Lin:2018cin} that allow to promote the 
search \cite{Sirunyan:2017dgc} for the gluon fusion $H\to b\bar{b}$ channel from just a search into the possibility of 
an observation.

For further details and resources I would like to refer to the references within the cited studies, as well as to a 
recent Standard Model working group report \cite{Bendavid:2018nar} and a recent review article on Higgs physics 
\cite{Dawson:2018dcd}.

\subsection*{Acknowledgments}

I would like to thank the organizers and participants of CIPANP 2018 for a well-organized and highly interesting 
conference at a great location.

This work was supported by the U.S.\ Department of Energy under award No.\ DE-SC0008347. This document was prepared 
using the resources of the Fermi National Accelerator Laboratory (Fermilab), a U.S. Department of Energy, Office of 
Science, HEP User Facility. Fermilab is managed by Fermi Research Alliance, LLC (FRA), acting under Contract No.\ 
DE-AC02-07CH11359.

\bibliographystyle{JHEP}
\bibliography{CIPANP2018-Neumann.bib}

\end{document}